
%
%
%
\newdimen\fullhsize \newdimen\hstitle \newdimen\hsbody
\tolerance=1000\hfuzz=2pt %

%
\magnification=1200\baselineskip=14truept plus 2truept minus 1truept
\hsize=6truein \vsize=8.5truein
\hsbody=\hsize \hstitle=\hsize 
%
%
\catcode`\@=11 
\newcount\yearltd\yearltd=\year\advance\yearltd by -1900
\def\Title#1#2{\nopagenumbers\abstractfont\hsize=\hstitle\rightline{#1}%
\vskip 1in\centerline{\titlefont #2}\abstractfont\vskip .5in\pageno=0}
\def\Date#1{\vfill\leftline{#1}\tenpoint\supereject\global\hsize=\hsbody%
}
%

\def\draftmode{\message{ DRAFTMODE }\def\draftdate{{\rm preliminary draft:
\number\month/\number\day/\number\yearltd\ \ \hourmin}}%
\headline={\hfil\draftdate}\writelabels\baselineskip=20pt plus 2pt minus 2pt
 {\count255=\time\divide\count255 by 60 \xdef\hourmin{\number\count255}
  \multiply\count255 by-60\advance\count255 by\time
  \xdef\hourmin{\hourmin:\ifnum\count255<10 0\fi\the\count255}}}
\def\nolabels{\def\wrlabel##1{}\def\eqlabel##1{}\def\reflabel##1{}}
\def\writelabels{\def\wrlabel##1{\leavevmode\vadjust{\rlap{\smash%
{\line{{\escapechar=` \hfill\rlap{\sevenrm\hskip.03in\string##1}}}}}}}%
\def\eqlabel##1{{\escapechar-1\rlap{\sevenrm\hskip.05in\string##1}}}%
\def\reflabel##1{\noexpand\llap{\noexpand\sevenrm\string\string\string##1}}}
\nolabels
%
\global\newcount\secno \global\secno=0
\global\newcount\meqno \global\meqno=1
\def\newsec#1{\global\advance\secno by1\message{(\the\secno. #1)}
\global\subsecno=0\xdef\secsym{\the\secno.}\global\meqno=1
\bigbreak\bigskip\noindent{\bf\the\secno. #1}\writetoca{{\secsym} {#1}}
\par\nobreak\medskip\nobreak}
\xdef\secsym{}
\global\newcount\subsecno \global\subsecno=0
\def\subsec#1{\global\advance\subsecno by1\message{(\secsym\the\subsecno. #1)}
\global\meqno=1 \ifnum\lastpenalty>9000\else\bigbreak\fi
\noindent{\it\secsym\the\subsecno #1}\writetoca{\string\quad
{\secsym\the\subsecno} {#1}}\par\nobreak\medskip\nobreak}
\def\appendix#1#2{\global\meqno=1\global\subsecno=0\xdef\secsym{\hbox{#1.}}
\bigbreak\bigskip\noindent{\bf Appendix #1. #2}\message{(#1. #2)}
\writetoca{Appendix {#1.} {#2}}\par\nobreak\medskip\nobreak}
%
%
\def\eqnn#1{\xdef #1{(\secsym\the\meqno)}\writedef{#1\leftbracket#1}%
\global\advance\meqno by1\wrlabel#1}
\def\eqna#1{\xdef #1##1{\hbox{$(\secsym\the\subsecno.\the\meqno##1)$}}
\writedef{#1\numbersign1\leftbracket#1{\numbersign1}}%
\global\advance\meqno by1\wrlabel{#1$\{\}$}}
\def\eqn#1#2{\xdef #1{(\secsym\the\subsecno.\the\meqno)}%
\writedef{#1\leftbracket#1}%
\global\advance\meqno by1$$#2\eqno#1\eqlabel#1$$}
%
%
\def\vfootnote#1{\insert\footins\bgroup\sevenpoint
\interlinepenalty=\interfootnotelinepenalty
\splittopskip=\ht\strutbox
\splitmaxdepth=\dp\strutbox \floatingpenalty=20000
\leftskip=0pt \rightskip=0pt \spaceskip=0pt \xspaceskip=0pt
\textindent{#1}\footstrut\futurelet\next\fo@t}
\newskip\footskip\footskip9pt 
\def\f@@t{\baselineskip\footskip\bgroup\aftergroup\@foot\let\next}
\setbox\strutbox=\hbox{\vrule height9.5pt depth4.5pt width0pt}
\global\newcount\ftno \global\ftno=0
\def\foot{\global\advance\ftno by1\footnote{$^{\the\ftno)}$}}
%
\newwrite\ftfile
\def\footend{\def\foot{\global\advance\ftno by1\chardef\wfile=\ftfile
$^{\the\ftno}$\ifnum\ftno=1\immediate\openout\ftfile=foots.tmp\fi%
\immediate\write\ftfile{\noexpand\smallskip%
\noexpand\item{f\the\ftno:\ }\pctsign}\findarg}%
\def\footatend{\vfill\eject\immediate\closeout\ftfile{\parindent=20pt
\centerline{\bf Footnotes}\nobreak\bigskip\input foots.tmp }}}
\def\footatend{}
%
%
\global\newcount\refno \global\refno=1
\newwrite\rfile
\def\up#1{$^{#1}$}
\def\ref{\up{\the\refno}\nref}
\def\nref#1{\xdef#1{\the\refno}\writedef{#1\leftbracket#1}%
\ifnum\refno=1\immediate\openout\rfile=refs.tmp\fi
\global\advance\refno by1\chardef\wfile=\rfile\immediate
\write\rfile{\noexpand\item{#1.\ }\reflabel{#1\hskip.31in}\pctsign}\findarg}
\def\findarg#1#{\begingroup\obeylines\newlinechar=`\^^M\pass@rg}
{\obeylines\gdef\pass@rg#1{\writ@line\relax #1^^M\hbox{}^^M}%
\gdef\writ@line#1^^M{\expandafter\toks0\expandafter{\striprel@x #1}%
\edef\next{\the\toks0}\ifx\next\em@rk\let\next=\endgroup\else\ifx\next\empty%
\else\immediate\write\wfile{\the\toks0}\fi\let\next=\writ@line\fi\next\relax}}
\def\striprel@x#1{} \def\em@rk{\hbox{}}

\def\addref#1{\immediate\write\rfile{\noexpand\item{}#1}} 
\def\startrefs#1{\immediate\openout\rfile=refs.tmp\refno=#1}
\def\xref{\expandafter\xr@f}\def\xr@f#1{#1}
\def\refs#1{\r@fs #1{\hbox{}}}
\def\r@fs#1{\edef\next{#1}\ifx\next\em@rk\def\next{}\else
\ifx\next#1\xref #1\else#1\fi\let\next=\r@fs\fi\next}
%

%
\newwrite\ffile\global\newcount\figno \global\figno=1
\def\fig{fig.~\the\figno\nfig}
\def\nfig#1{\xdef#1{fig.~\the\figno}%
\writedef{#1\leftbracket fig.\noexpand~\the\figno}%
\ifnum\figno=1\immediate\openout\ffile=figs.tmp\fi\chardef\wfile=\ffile%
\immediate\write\ffile{\noexpand\medskip\noexpand\item{Fig.\ \the\figno. }
\reflabel{#1\hskip.55in}\pctsign}\global\advance\figno by1\findarg}
\def\xfig{\expandafter\xf@g}\def\xf@g fig.\penalty\@M\ {}
\def\figs#1{figs.~\f@gs #1{\hbox{}}}
\def\f@gs#1{\edef\next{#1}\ifx\next\em@rk\def\next{}\else
\ifx\next#1\xfig #1\else#1\fi\let\next=\f@gs\fi\next}
\newwrite\lfile
{\escapechar-1\xdef\pctsign{\string\%}\xdef\leftbracket{\string\{}
\xdef\rightbracket{\string\}}\xdef\numbersign{\string\#}}

\def\writestop{\def\writestoppt{\immediate\write\lfile{\string\pageno%
\the\pageno\string\startrefs\leftbracket\the\refno\rightbracket%
\string\def\string\secsym\leftbracket\secsym\rightbracket%
\string\secno\the\secno\string\meqno\the\meqno}\immediate\closeout\lfile}}
\def\writestoppt{}\def\writedef#1{}
\def\seclab#1{\xdef #1{\the\secno}\writedef{#1\leftbracket#1}\wrlabel{#1=#1}}
\def\subseclab#1{\xdef #1{\secsym\the\subsecno}%
\writedef{#1\leftbracket#1}\wrlabel{#1=#1}}
\newwrite\tfile \def\writetoca#1{}
\def\leaderfill{\leaders\hbox to 1em{\hss.\hss}\hfill}
\def\writetoc{\immediate\openout\tfile=toc.tmp
   \def\writetoca##1{{\edef\next{\write\tfile{\noindent ##1
   \string\leaderfill {\noexpand\number\pageno} \par}}\next}}}

%
\catcode`\@=12 
%

\font\titlerm=cmr10 scaled\magstep3 \font\titlerms=cmr7 scaled\magstep3
\font\titlermss=cmr5 scaled\magstep3 \font\titlei=cmmi10 scaled\magstep3
\font\titleis=cmmi7 scaled\magstep3 \font\titleiss=cmmi5 scaled\magstep3
\font\titlesy=cmsy10 scaled\magstep3 \font\titlesys=cmsy7 scaled\magstep3
\font\titlesyss=cmsy5 scaled\magstep3 \font\titleit=cmti10 scaled\magstep3
\skewchar\titlesy='60 \skewchar\titlesys='60 \skewchar\titlesyss='60
\def\titlefont{\def\rm{\fam0\titlerm}
\textfont0=\titlerm \scriptfont0=\titlerms \scriptscriptfont0=\titlermss
\textfont1=\titlei \scriptfont1=\titleis \scriptscriptfont1=\titleiss
\textfont2=\titlesy \scriptfont2=\titlesys \scriptscriptfont2=\titlesyss
\textfont\itfam=\titleit \def\it{\fam\itfam\titleit} \rm}
\def\abstractfont{\tenpoint}
\def\tenpoint{\def\rm{\fam0\tenrm}
\textfont0=\tenrm \scriptfont0=\sevenrm \scriptscriptfont0=\fiverm
\textfont1=\teni  \scriptfont1=\seveni  \scriptscriptfont1=\fivei
\textfont2=\tensy \scriptfont2=\sevensy \scriptscriptfont2=\fivesy
\textfont\itfam=\tenit \def\it{\fam\itfam\tenit}
\textfont\bffam=\tenbf \def\bf{\fam\bffam\tenbf} \rm}
\font\sevenrm=cmr7 scaled \magstephalf
 \font\seveni=cmmi7 scaled \magstephalf
\font\sevensy=cmsy7 scaled \magstephalf
\font\sevenbf=cmbx7 scaled \magstephalf
\font\sevenit=cmti7 scaled \magstephalf

\def\sevenpoint{\def\rm{\fam0\sevenrm}
\textfont0=\sevenrm \scriptfont0=\fiverm
\scriptscriptfont0=\fiverm
\textfont1=\seveni  \scriptfont1=\fivei  \scriptscriptfont1=\fivei
\textfont2=\sevensy \scriptfont2=\fivesy \scriptscriptfont2=\fivesy
\textfont\itfam=\sevenit \def\it{\fam\itfam\sevenit}
\textfont\bffam=\sevenbf \def\bf{\fam\bffam\sevenbf} \rm}
%
%
%

\hyphenation{anom-aly anom-alies coun-ter-term coun-ter-terms}
\def\inv{^{\raise.15ex\hbox{${\scriptscriptstyle -}$}\kern-.05em 1}}

\def\Dsl{\,\raise.15ex\hbox{/}\mkern-13.5mu D} 
\def\dsl{\raise.15ex\hbox{/}\kern-.57em\partial}


\def\boxeqn#1{\vcenter{\vbox{\hrule\hbox{\vrule\kern3pt\vbox{\kern3pt
	\hbox{${\displaystyle #1}$}\kern3pt}\kern3pt\vrule}\hrule}}}
\def\mbox#1#2{\vcenter{\hrule \hbox{\vrule height#2in
		\kern#1in \vrule} \hrule}}  
%

\def\darr#1{\raise1.5ex\hbox{$\leftrightarrow$}\mkern-16.5mu #1}

\def\roughly#1{\raise.3ex\hbox{$#1$\kern-.75em\lower1ex\hbox{$\sim$}}}


\Title{SSCL--PP-158}{\vbox{\hbox{\centerline{The SSC:}}
\hbox{ Programme and Searches for New Particles}}}

\centerline{Benjam\'\i n Grinstein\footnote{$^\dagger$}
{email: grinstein@sscvx1.bitnet, \ @sscvx1.ssc.gov}}
\bigskip\centerline{Superconducting Super Collider Laboratory}
\centerline{2550 Beckleymeade Ave, Dallas, Texas 75237}


\vskip .3in
Talk presented at the Trieste Workshop on the Search for New Elementary
Particles: Status and Prospects.

\Date{10/92} 


\centerline{{\bf THE SSC: PROGRAMME AND SEARCHES FOR NEW PARTICLES}}
\vskip0.5truein
\centerline{BENJAMIN GRINSTEIN}
{\sl \centerline{Superconducting Super Collider Laboratory}
\centerline{2550 Beckleymeade Ave, Dallas, Texas 75237}}

\newsec{   Introduction}

The Superconducting Super Collider (SSC) Laboratory is under
construction.  Many critical components have been designed, small
and full scale prototypes built and successfully tested, and in some
cases samples have been produced by industry. The `Project', as we
call the task of building the SSC Lab, is well under way.  In this
talk I will describe the Project and give an idea of what its status
is to date.

The physical potential of the SSC is extraordinary.  Whatever is
responsible for the spontaneous breaking of the electroweak
symmetry, it must loom at the energies accessible at the SSC.
It will produce top quarks copiously.  And, most
significantly, it will probe the shortest distance scales ever,
opening up a new frontier in High Energy Physics.  Also in this
talk, I will touch on some of the physics of the SSC.

I end the introduction with a disclaimer. This talk is about subjects outside
my field of expertise. Describing the Project would  be better left to an
administrator or manager, while a discussion of what could be found and how,
could certainly be done more properly by  one of the many brilliant
experimenters involved in the design of the major detectors for the SSC. Not
all is lost, though. You will get what I hope is an interesting and peculiar,
if not illuminating, insider/outsider's view. Insider's because I am employed
by the SSC, and so I am  embedded in the day to day discussions about the
Project. Outsider's because I am neither a manager nor an experimentalist.

Most of the material covered in this talk comes from conversations
 with my SSC colleagues and from their transparencies. I have
 also used some technical documents.\nref\sdc{Technical Design Report,
 Solenoidal Detector Collaboration, 1 April 1992, SDC-92-201}%
\nref\gem{Letter of Intent, GEM Collaboration, November 30, 1991,
 GEM-92-49}\up{\refs{\sdc,\gem}}

\newsec{  The Project}

\subsec{ The accelerator systems}

The SSC is a facility for producing two counter-rotating high
energy and high intensity proton-proton beams, and for studying
their collisions. The design of the facility was made on the premise
that interesting physics would be accessible to the laboratory if the
center of mass energy were to be at least 20 TeV, and that integrated
luminosities of  $10^{40}{\rm{cm}}^{-2}$ would be needed to make
discoveries possible.  The design parameters of each of the two
collider rings were chosen to meet these criteria; see table~1.


\vskip0.8cm

\centerline{Table 1. SSC Parameters}

{\settabs5\columns
\+&Proton Energy						&&20 Tev\cr
\+&Circumference of Rings &&     87 Km\cr
\+&Protons per r.f. bunch			&& 0.75$\times10^{10}$\cr
\+&Bunch Spacing        &&    5 m\cr
\+&   Number of Bunches(per ring)      &&    17,424 \cr
\+&   Total Particle Energy (per ring)    &&  418$\times 10^6$ J\cr
\+&Emittance (RMS) &&  $1\pi$ mm-mrad\cr
\+&Interaction Region Focal Spot &&5$\times10^{-6}$ m\cr
\+&~~~~ Size RMS Radius
($\beta^*=0.5$m)\cr
\+&$p$-$p$ collisin rate    &&   60 MHz\cr
\+&Luminosity    &&      $1\times10^{33}$ cm$^{-2}$ sec$^{-1}$\cr
\+&Synchrotron Radiation Power   &&   8.75 KW (per ring)\cr
 }
\vskip0.6cm

Four stages of injection will accelerate protons from zero to 2 TeV,
for delivery  into the collider rings.  The parameters of these
accelerators are given in table~2.

The first stage, the linac, is really an array of several systems
that work in series. An Ion Source is followed by the Low Energy
Beam Transport.  The 2.2m long Radio Frequency Quadrupole
accelerates the H$^-$s from 35~keV to 2.5~MeV, which are then
accelerated to 70~MeV in the 23m Drift Tube Linac. The last system of
the linac is the 120m Coupled Cavity Linac which boosts the ions to
an energy of 600~MeV. Three days before  this talk was given news
came
out of a \$4.9 million general construction contract awarded to
Sedalco, Inc. of Fort Worth, Texas, to begin construction of the
facilities that  will house the Linac.



\vskip0.8cm
\centerline{Table 2. Injector and Collider Parameters.}
{\tabskip=1em plus 3em minus 0.5em
\halign to\hsize{#\hfil&\hfil#&\hfil#&\hfil#&\hfil#&\hfil#\cr
&Linac&LEB&MEB&HEB&Collider\cr
Kinetic Energy (GeV) & 0.6& 11.1& 180--200& 2,000& 20,000\cr
Circumference (Km)&   (length)0.11&0.54&3.96&10.89&87.12\cr
Cycle Time&0.1sec&0.1sec&3sec&2min&$\sim$24 hr\cr
Protons per Cycle& --- & $10^{12}$& $8\times10^{12}$& $2\times10^{13}$&
$1.3\times10^{14}$\cr
Protons per Bunch& --- & $10^{10}$& $10^{10}$& $10^{10}$&
$0.75\times10^{10}$\cr
Normalized RMS Emittance& $<0.5$& 0.6& 0.7&0.8& 1.0\cr
{}~~~~~($\pi$ mm-mrad)\cr
}
}
\vskip0.6cm

The Low and Medium Energy Boosters use warm magnet systems,
while the
High Energy Booster and the Collider use cold magnets. The design of
the
High Energy Booster is both interesting and challenging: to provide
protons to the two collider rings it must handle rotating and
counter-rotating beams, and this is accomplished by quickly
alternating
the polarity of the  dipoles.

It is interesting that the location chosen for the collider has
rather favorable geological attributes. Most of the collider tunnel
will be bore through a layer of Austin Chalk, a type of limestone.
Below this lays some Eagle Ford Shale. My knowledge of geology is
very limited; my son reads to me from his book that `limestone is
made of shells all pressed together, sandstone is made of grains of
sand all pressed together, shale is made of mud and clay all pressed
together'. The point is that,  under pressure, shale behaves like mud
but limestone is more like concrete. Earlier this year a decision
was made to locate the experimental halls for the large multipurpose
detectors (see below) on the east side. With the main campus on the
west side, this is certain to make it inconvenient to the
users of those east side facilities. The reason for the decision is
that experimental halls for large detectors on the west side would
reach down into the shale, making it hard to stabilize a very
heavy detector.

\subsec{ Magnet Systems}

The success of this enterprise lies heavily on the ability to
produce industrially high quality superconducting dipoles. So not
only must dipoles be designed and constructed in laboratories that
will conform to demanding specifications, but the technology has to
be transfered to industry.

Table~3 displays the characteristics of the collider's dipoles. The
performance of dipoles constructed both by FNAL and BNL, and by
technicians from industry working at these labs is superb. Typical
quench sequences (at 4.35$^\circ$K) involve at most one quench below
the
operating current of 6500 Amps (on the first ramp up), with every
subsequent quench occurring always at currents exceeding
7000~Amps.



\magnification=\magstep1

\vskip0.8cm

\centerline{Table 3. Parameters of SSC Superconducting Dipoles (and
Quadrupoles)}

{\settabs5\columns
\+&Operating Field           &&       6.6 T\cr
\+&(Operating Gradient)   &&   206  T/m\cr
\+&Operating Current   && 6.5 kAmps    \cr
\+&Stored Energy    && 1.58 MJ (1.32 MJ)    \cr
\+&Length      &&  15.8 m (13.3 m)   \cr
\+&Inner Coil Aperture      &&50 mm     \cr
\+&Cold Mass      &&12,700 kg (10,800 kg)     \cr
\+& Conductor     && NbTi\cr
\+& $J_c$ (5T, 4.2$^\circ$K)     && 2750 Amps/mm     \cr
\+&Filament Diameter      &&  6 $\mu$m   \cr
\+&  Operating Temperature    &&  4.35$^\circ$K   \cr
\+&  Dipoles (Quads) per Ring    &&  4230 (832)  \cr
}
\vskip0.6cm

The Accelerator Systems String Test (ASST) has as objective  operating a half
cell of five collider dipole magnets, one quadrupole magnet, and two spool
pieces at the design current of 6500 Amps. It is housed in a long structure
built to reproduce precisely the (almost imperceptible) actual curvature of
the tunnel. A half cell is the minimum reproducible unit of which the collider
arcs are built. Meeting the AAST objectives successfully is necessary to
demonstrate that the many different systems that go  into a minimal collider
unit can work together.

\subsec{ Other systems and facilities. Schedule}

The SSC Laboratory will be a complex array of very many systems. It
is impossible to properly describe them here, so I won't. Just as an
example of what is involved, the cryogenic systems for the collider
will collectively perform as  the biggest cooling plant in the world.
There are also satellite facilities with important functions. For
example, in collaboration with Southwest Medical Center a medical
applications station has been planned. It will use H$^-$ beams from
the linac for cancer therapy, among other applications.



\vskip0.8cm
\centerline{Table 4. SSC Baseline Schedule: Major Project Milestones.}
{\tabskip=1em plus 3em minus 0.5em
\halign to\hsize{#\hfil&\hfil#\cr
\hfil \it \underbar{Description} & \it \underbar{Schedule}  \cr
Baseline Validation Complete & July 1990\cr
A-E/CM Letter Contract \& NTP & August 1990\cr
CDM Authorization to Incur Costs & November 1990\cr
SEIS Record of Decision (ROD) & February 1991\cr
Begin Conceptual Design for Detectors & February 1991\cr
Start SSC Civil Construction & February 1991\cr
\bf Accelerator String Test Complete & \bf September 1992\cr
Notice to Proceed Experiment Halls & June 1993\cr
Full-rate Production Decision on Magnets & April 1994\cr
Start First Half Sector -- CDM Delivery &  April 1994\cr
First Collider Half Sector -- Start Installation & June 1994\cr
LINAC Start Commissioning (600 MeV) & October 1994\cr
First Collider Half Sector -- Start Cooldown & March 1995\cr
LEB Start Commissioning & October 1995\cr
Beneficial Occupancy of Large Experimental Halls & January 1996\cr
MEB  Start Commissioning & June 1996\cr
HEB Start Installation & August 1996\cr
MEB Test Beams Available & January 1997\cr
HEB Start Commissioning & September 1998\cr
West Detectors -- Start Commissioning & March 1999\cr
Collider -- Start Commissioning & March 1999\cr
Beam to Experiments (End of Project/Begin Ops.) & September 1999\cr
}
}
\vskip0.6cm

At the beginning of time (1990) a document, the `SSC Baseline' was
created.
(As far as the Project is concerned, the
beginning of time {\it is} the baseline). In varying degrees of
detail, the baseline specs out the project, its timetable and cost.
Table~4 lists the major project milestones together with their
baseline schedule. Of note is that the Accelerator Systems String
Test (see 2.2, above) met its objectives six weeks ahead of schedule
(at 11:39 AM, on Friday, August 14). The cooldown for the key test
began
in June and the current was slowly raised in the string, testing the
quench protection system along the way.  The test demonstrated the
quality of the industrially assembled magnets and the associated
power,
cooling, and control equipment to operate together successfully as a
system.

Also of note, beam should be delivered to experiment by September
1999.
Of course,  it will take some additional time before the beam to
experiment achieves design luminosity. Studies of operation
procedures
are currently under way.

\newsec{  Experimental Program}

\subsec{ Generalities}

The initial experimental program of the SSC should match the physics
potential and the investment in such a discovery machine.  It was
therefore decided,  upon recommendation of the advisory panels,
that the initial program should involve:
\item{$\triangleright$}Two major, complementary detectors aimed at the physics
opportunities opened at multi-TeV energies; and
\item{$\triangleright$}Several smaller experiments, adding diversity and
coverage  of phenomena at low transverse momentum and B-physics.

The  experimental program of the SSC is proposal driven. Expressions
of Interest were called for starting in May 1990. By the time this talk
was given 21 Expressions of Interest had been submitted, involving
some 2300 collaborators from about 350 institutions worldwide.
They include large all purpose detectors and smaller specialized
detectors. The latter include, among others,  a proposal for study of
``Low $p_T$ Physics at the SSC'', four for studies of $B$ decays (and
CP violation), one proposal for ``A Very Long Baseline Neutrino
Oscillation Experiment'' and one for ``A Full Acceptance Detector for
SSC Physics at Low and Intermediate Mass Scales''.

(If I understand correctly) Decisions on the experimental program
are made by the Director of the Laboratory upon recommendation
from advisory panels (the Scientific Policy Committee and the
Program Advisory Committee), constituted from renowned
international experts.

\subsec{ Status of the Two Major All Purpose Detectors}

The Program Advisory Committee stated (July 1990)  that ``A
healthy initial program requires two detectors with complementary
as well as overlapping strengths that address the physics at high
$p_T$''.  That their strengths be complimentary has the obvious
effect of increasing the total capabilities and discovery reach.
Overlapping strengths allow for cross checks of discoveries. And, of
course, that there are two experiments with some overlapping
capabilities introduces that element of competitiveness that  brings the
best out of scientists!

Out of the many Expressions of Interest for major detectors
eventually two were selected to proceed forward with Letters of
Intent and then Technical Design Reports:

\itemitem{GEM}Its design Goals emphasize identification and precision
measurement of gammas, electrons and muons, with capabilities for
higher luminosity. The Letter of Intent was submitted on November
29, 1991, and in January 1992 was given approval to proceed
towards a Technical Design Report. At present detector R \& D and
Engineering is under way, with some major decisions for
calorimetry to be made by the fall of 1992.
\itemitem{SDC}Its design Goals emphasize charged particle tracking,
hermetic calorimetry, lepton energy measurement and identification
and vertex detection. Letter of Intent was submitted on November
30, 1990, and in January 1991 was given approval to proceed
towards a Technical Design Report. This was submitted on April 1, 1992.
Detector R \& D and Engineering is under way.

\subsec{ Physics}

As an example of the physics that can be studied at the SSC, we focus
on
discovery of a standard model higgs. By 1999 LEPII will have either
discovered the higgs or set a lower bound on its mass of about
80~GeV.
Theoretical upper bounds on the standard model higgs mass are in
the
600~GeV to 800~GeV range. Therefore a strategy for higgs discovery at
SSC
needs only cover the 80~GeV to 800~GeV mass range.

The main problem for almost any SSC experiment is the enormous
two jet
cross section. This translates into large, often unmanageable detector
backgrounds. For example, the rate for higgs production  anywhere
in the
above mass range is at least five orders of magnitude smaller than the
rate  of
two jet events. Thus discovery in (or study of) purely hadronic decay
channels, such as $H\to b\bar b$ is nearly impossible.

Discovery strategies are driven by the higgs decay fractions. For
masses
above $2M_Z$ the dominant decay modes are into pairs of weak
vector
bosons. The cleanest channel is ${\rm higgs}\to ZZ\to\ell^+\ell^-\ell^+\ell^-$,
but it is statistically limited. Not as clean is
${\rm higgs}\to ZZ\to\ell^+\ell^-\nu\bar\nu $. Much more abundant are the
processes  ${\rm higgs}\to ZZ\to\ell^+\ell^-+{\rm 2 jets}$ and
 ${\rm higgs}\to W^+W^-\to\ell\nu+{\rm 2 jets}$, which are respectively
20 and 150 times more frequent than the four charged leptons
mode. They are less clean,
as they involve backgrounds from $W/Z+{\rm jets}$ and $t\bar t$, so in the end
they give smaller signal to background ratios. The main backgrounds to
${\rm higgs}\to ZZ\to\ell^+\ell^-\ell^+\ell^-$ are from $q\bar q$ or $gg\to
ZZ$,
$Z+Q\bar Q$ or $Q\bar Q$. An additional complication comes from the fact that
the width of the higgs boson increases from 1.4 to 30 to 270 GeV as the mass
increases from 200 to 400 to 800 GeV. Except at the upper end of this mass
range, simple cuts seem to ensure discovery of the higgs boson in the four
lepton channel in 1 SSC year (1 year
of running at design luminosity, i.e., $10^{40} {\rm cm}^{-2}$). At the upper
end of the mass range either the inclusion of other channels, or additional
integrated luminosity, or both, would be necessary for discovery.

Below $2M_Z$ decay into two bottom quarks quickly becomes
dominant, but the decay into $ZZ^*$, i.e., into $Zl^+l^-$, $Z\nu\bar\nu$ and
$Zq\bar q$, is still significant and can be used for discovery in the range
120~GeV $\le M_h < 2M_Z$. (The branching fraction into $ZZ^*$ dips to a
minimum at $\sim2M_Z$, reaching a maximum at $\sim150~{\rm GeV}$). The most
important backgrounds for  ${\rm higgs}\to ZZ^*\to e^+e^-e^+e^-$  come from
production of $t\bar t$, $Zb\bar b$ and $Zt\bar t$; it seems possible to bring
these under control through simple transverse momentum and rapidity  cuts. The
situation is similar for other channels. The combined signal for $4e$, $4\mu$
and $2e2\mu$ seems very significant after 1 SSC year.

In the lowest end of the mass range, $80~ {\rm GeV} <
M_h < 130~ {\rm GeV}$,  the  decay
mode of choice is into two photons. Although its branching fraction is
small (approximately constant and $\sim10^{-3}$ in the range 80 Gev$< M_h<150$
Gev), it is a relatively clean mode. Directly produced higgs with subsequent
decay into two photons seems hard, if not impossible, at SDC and GEM; the
irreducible backgrounds from $q\bar q\to\gamma\gamma$ and $gg\to\gamma\gamma$
are several orders of magnitude larger than the signal. If the higgs is
produced in association with another particle which may be used to tag the
event the prospects of discovery seem brighter. Studies by both SDC and GEM
indicate that the combined $W+{\rm higgs}$ and $t\bar t+{\rm higgs}$ signals
may yield a 4-sigma discovery limit in this mass range within 1 SSC year. The
mass resolution in
this range clearly depends critically on the specific choices made for
electromagnetic calorimetry, and is estimated to be somewhere between 0.5~GeV
and  2.5~GeV.

Space limitations preclude us from discussing the many other discovery
potentials of the SSC experiments. The list is long (Supersymmetric partners of
standard particles, technicolor resonances, new gauge bosons, multiple higsses,
and on and on), {\sl but hopefully incomplete!\/} What makes the SSC so unique
among present day accelerators is that it is designed to probe the dynamics
underlying the breaking of the electroweak symmetry, the stuff of which matter
is made. The mechanism responsible for this could be  completely new and
unanticipated!

\newsec{  Concluding Remarks}

The odyssey has begun.  Tunnels are being dug, magnets are
being fabricated, systems are being tested. Two gargantuan experimental
collaborations have been formed and are quickly  transiting from the design
 into the  development stage. The SSC counts with over 2200 employees, and
\$517 million in funding for Fiscal Year 1993 has been signed into law. If
everything proceeds according to plan (and so far it has), we may very well
have a first look at a completely new repertoire of high energy phenomena by
2001.

\footatend
\immediate\closeout\rfile\writestoppt
\baselineskip=14pt
\vskip1truecm plus 0.4truecm minus 0.2truecm
\newsec{ References}
\bigskip{\frenchspacing%
\parindent=20pt
\escapechar=` \input refs.tmp\vfill\eject}\nonfrenchspacing
\bye